\pdfoutput=1
\documentclass[10pt, conference, compsocconf]{IEEEtran}
\usepackage{cite}
\usepackage[cmex10]{amsmath}
\usepackage{algorithmic}
\usepackage{array}
\usepackage[pdftex]{graphicx}
\usepackage{textcomp}
\usepackage{latexsym}
\usepackage{mathptmx}
\usepackage{fixltx2e}
\RequirePackage{accsupp}
\usepackage{pdfcomment}
\usepackage{comment}
\usepackage[autostyle]{csquotes}
\def\BibTeX{{\rm B\kern-.05em{\sc i\kern-.025em b}\kern-.08em
    T\kern-.1667em\lower.7ex\hbox{E}\kern-.125emX}}

\begin{document}

\title{This One Simple Trick Disrupts Digital Communities}

\author{\IEEEauthorblockN{Philip Feldman}
\IEEEauthorblockA{\textit{ASRC Federal} \\
Columbia, USA \\
philip.feldman@asrcfederal.com}
\and
\IEEEauthorblockN{Aaron Dant}
\IEEEauthorblockA{\textit{ASRC Federal} \\
Columbia, USA \\
aaron.dant@asrcfederal.com}
\and
\IEEEauthorblockN{Wayne Lutters}
\IEEEauthorblockA{\textit{Dept. of Human Centered Computing} \\
\textit{University of MD, Baltimore County}\\
Catonsville, USA \\
lutters@umbc.edu}
}

\maketitle

\begin{abstract}
This paper describes an agent based simulation used to model human actions in belief space, a high-dimensional subset of  information space associated with opinions. Using insights from animal collective behavior, we are able to simulate and identify behavior patterns that are similar to nomadic, flocking and stampeding patterns of animal groups. These behaviors have analogous manifestations in human interaction, emerging as solitary explorers, the fashion-conscious, and echo chambers, whose members are only aware of each other. We demonstrate that a small portion of nomadic agents that widely traverse belief space can disrupt a larger population of stampeding agents. We then model the concept of Adversarial Herding, where trolls, adversaries or other bad actors can exploit properties of technologically mediated communication to artificially create self sustaining runaway polarization. We call this condition the \textit{Pishkin Effect} as it recalls the large scale buffalo stampedes that could be created by native Americans hunters. We then discuss opportunities for system design that could leverage the ability to recognize these negative patterns, and discuss affordances that may disrupt the formation of natural and deliberate echo chambers. 
\end{abstract}

\begin{IEEEkeywords}
belief space, computer simulation, echo chamber, flocking,  group processes, polarization, social behavior
\end{IEEEkeywords}

\section{Introduction}
Emergent collective behavior such as flocking, schooling and swarming is a curiously universal phenomena. In addition to animal behavior, social influence has been studied in such diverse contexts as coupled oscillators, consensus formation in networks, load balancing, and belief propagation \cite{olfati2007consensus}. Collective behavior based on social influence clearly has benefits. Being in a group offers protection from predators and can be helpful in finding food, particularly in cases where traces are faint \cite{grunbaum1998schooling}. Human collective actions extends these behaviors into information domains, where they can manifest in the writing of government constitutions \cite{rutherford2016disentangling}, revolutions \cite{lynch2017online}, as well as more trivial, but socially important examples such as fashion \cite{curran1999analysis}. 

Collective behavior exists on a continuum. At one end, nomadic patterns have individuals or small groups dispersed over large areas. Though these nomads do not enjoy the protection of large groups, the overall population gains resilience due to its extended footprint. Further along the spectrum are various types of clustering - flocks, herds and schools. This higher level of social interaction can increase overall information processing leading to better sensing and more optimal exploitation of resources \cite{grunbaum1998schooling}. At the far end of the spectrum, social influence becomes the dominant factor, outweighing such things as environmental cues. These conditions lead to social inertia, where the collective is unable to adapt to changing conditions. In the extreme, this can lead to panics and stampedes \cite{torney2015social}. Often, to balance out nomadic costs and the risks of social inertia, a population will incorporate an explore/exploit strategy \cite{berger2014exploration}, where a small portion of the population will range beyond the traditional habitat. This strategy provides a level of resilience for the overall population while letting the majority exploit opportunities discovered by the nomads. 

Stampedes by animals, a mass movement in a unified direction \cite{ward1987cowboy} is analogous to echo chambers or  groupthink, a form of extreme cohesion that creates its own social reality \cite{burnette2011leadership}. The most dramatic case of this may be Nazi Germany, where Arendt wrote of \enquote{[a] movement which is being propelled with increasing speed in a certain direction} \cite{arendt1973origins}. Moscovici and Doise\cite{moscovici1994conflict} describe how groups in free discussion compile complex information into simplified norms as a mechanism to achieve agreement and polarization. Norms are the direction that the group aligns with, and have a relationship to the group's coherence and rate of change. Munson and Resnik \cite{munson2010presenting} show online user behavior (confirming, diversity-seeking, diversity-avoiding) that provides a computer-mediated framework these user actions.

When information is mediated through technology, the information horizons that inform where to \enquote{explore} and when to \enquote{exploit} can become obscured, leading to destructive behaviors. Technology makes available abundant information with wildly varying levels of veracity, structure, bias, and credibility, blurring the lines between naive belief and well-supported evidence. Beliefs are not facts, but they are \textit{factive} -- they \textit{feel} like facts \cite{denicola2017understanding}. This makes prima facie determination of the quality of specific information extremely difficult for humans and systems that depend on them\cite{lukoianova2014veracity}. Like stampedes in the animal kingdom, echo chambers, the self-reinforcing reflection of a belief \cite{key1966responsible}, are a symptom of this misalignment of environmental awareness and social influence.  In many cases, credible-looking, low-quality or misleading information contributes to the formation of echo chambers \cite{jafarinaimi2015collective}. Members often believe that they have access to all needed information and are unaware of critical perspectives \cite{flaxman2016filter}.

Information retrieval (IR) systems often sidestep the issue of information quality by tailoring results to user preferences, thus trusting in users' abilities to discern fact from fiction or opinion \cite{nied2017alternative}. Yet human assessments are vulnerable to cognitive biases, social identities, perceived pressure, norms, and other factors that interfere with accurate judgment \cite{ahmad2011interpreting}. The process of seeking information can lead to a self-confirmatory feedback loop, in which low quality information is perceived as valid and higher quality information is excluded until it closes off into an echo chamber. Once sealed in an internal discourse, the echo chamber can gain velocity and become a stampede. Our information technologies provide many tools to aid this process. Framed in a supporting context with a believable interface, large groups of people can be persuaded of many things - that a president is secretly from another country \cite{warner2014echoes} or that aliens have landed in New Jersey \cite{campbell2010getting}. 

The ability for IR-mediated group interactions to connect like-minded people who can't easily see out of their filter bubbles can lead to events such as the \enquote{Pizzagate conspiracy}, an echo chamber based on group belief that there was high ranking Democrat involvement in human trafficking \cite{aisch2016dissecting}. This echo chamber produced extremely dangerous real world outcomes \cite{jackson2017conspiracytheories}. Some solutions to this issue have been tried successfully. Chandrasekharan et al have shown that banning Reddit hate groups can disrupt echo chambers with continuing positive effects on post-membership users as they are re-integrated into a larger, more diverse community \cite{Chandrasekharan2018Reddit}. Salganik has shown that ranking that prioritizes quality over popularity \cite{salganik2008leading} can eliminate runaway effects in music selection. It seems reasonable that design changes at social websites like Facebook, Twitter or Reddit, to enhance information diversity \cite{thudt2012bohemian} could decrease the likelihood of such a runaway result.

To take all these separate pieces and understand how they function together, we need a model that can represent, in a simplified fashion, critical aspects of computer mediated group interaction. For this, we turn to agent-based simulation. Agent-based and cellular automata-based simulation has proven to be a particularly effective mechanism for modeling the complex interplay between individuals in a population. In problems ranging from neighborhood segregation\cite{schelling1971dynamic} to opinion dynamics\cite{hegselmann2002opinion} to culture dissemination\cite{sen2013sociophysics}, these types of simulations have been shown to be effective in generating complex emergent collective behaviors from sets of simple, understandable rules applied to the agent. 

\section{Previous Work}

Animal models have often served as a starting point for understanding human interaction with information. Danchen et. al. showed that animals and humans both use inadvertent social information (ISI) to influence decisions about environmental quality and appropriateness \cite{danchin2004public}. Card and Pirolli \cite{pirolli1999information} successfully demonstrated  the utility of animal models for individual human information foraging behaviors. Deneubourg and Goss' \cite{deneubourg1989collective} work related to animal group cognitive behaviors such as flocks and herds. More recently, Olfati--Saber et. al. have shown that social influence leading to collective behaviors is a widespread phenomenon in natural and artificial systems \cite{olfati2007consensus}. Connecting animal models to technology-mediated human group interaction, Belz et. al. have shown the emergence of spontaneous flocking in computer mediated communication \cite{belz2013spontaneous}.

The study of human group behavior has roots in the 19th-century work of LeBon \cite{le1897crowd}, who showed that crowds can move and think like single organisms, which was later studied experimentally by Moscovici \cite{moscovici1994conflict}.  More recently, Krause \cite{hegselmann2002opinion} and Bikhchandani \cite{bikhchandani1992theory} have modeled opinion dynamics and echo chambers while Salganik \cite{salganik2008leading} has demonstrated that online rating based on popularity can produce runaway results. Epstien et. al shows similar results for Information retrieval with the Search Engine Manipulation Effect \cite{epstein2015search}. 

Game theory has also explored this problem space, particularly with respect to the evolution of cooperation. Using the \textit{Iterated Prisoner's Dilemma}, research by Nowak and others show that there is dominant transition pattern that establishes from an initial random population. The first population is the antisocial \textit{Always Defect} (AllD), which transitions to more social \textit{Tit-for-Tat} (TFT), and continues through \textit{generous TFT} to the highly social, efficient, and vulnerable \textit{Always Cooperate} which in turn can be decimated by AllD \cite{nowak1992tit}. A more stable strategy is \textit{win-stay, lose-shift} \cite{nowak1993strategy}, where profitable strategies are maintained until they fail, at which point the agent explores other tactics. This resonates with aspects of the \textit{Multi-Armed Bandit Problem}, which examines when an agent should exploit a current slot machine, or explore for a better one \cite{gittins2011multi}. Bacharach \cite{bacharach2006beyond} explored a theoretical explanations for the above behavior which he extended to coordination games such as Stag Hunt. His insight was that when humans identify with groups, they can reason from that perspective and rationally choose options such as cooperate. 

Lastly, Curran \cite{curran1999analysis} shows qualities such as fads in fashion (a form of flocking behavior) can also be represented in a spatial way using movement in a belief space by observing norms for skirt length and width. She shows a detailed example of collective cognitive movement through a belief space traced over 36 years.

These models are effective and compelling descriptions, but are not optimal for producing emergent patterns. Bonabeau \cite{bonabeau2002agent} states that, \enquote{By definition, [emergent phenomena] cannot be reduced to the system's parts: the whole is more than the sum of its parts because of the interactions between the parts.} Reynolds\cite{reynolds1987flocks}, Cucker \cite{cucker2007emergent}, Olfati--Saber \cite{olfati2006flocking} and others have built and described agent-based simulations that produce emergent flocking, schooling, and herding characteristics that closely mimic observed animal behavior.

\section{Model Considerations}

Our base model explores two ideas: 1) that human navigation through \textit{belief space} (a subset of information space associated with opinions) is analogous to animal motion through physical space, and 2) that the \textit{digital inadvertent social information} (DISI) provided by humans interacting with the belief environment can be used to characterize the underlying space.

We rest this assumption on recent work on neural coupling \cite{stephens2010speaker} which has been used to show that social cognition is a \textit{physical} process where individual minds synchronize neural firing patterns to support mutual cognition \cite{gallotti2017alignment}. Since Olfati--Saber et. al. have shown that the underlying mathematics of collective behavior are shared across a wide range of constrained and unconstrained consensus domains \cite{olfati2007consensus}, we believe that mechanisms for navigating physical space can be extended to handle cognitive spaces. After extensive model development, we determined the minimum set of features that an agent requires re:

\begin{enumerate}
	\item \textit{Dimension} -- The number of beliefs that a person may hold is not limited by physical space. They may hold opinions on many subjects. To keep calculations manageable, we set an upper bound of 10 dimensions.
	\item \textit{Velocity} -- Humans and animals dynamically interact with their physical and information environments. Although they may have regions or territories that they prefer, movement in the physical and political sense is a defining characteristic.
	\item \textit{Heading} -- There appears to be a rate-limited alignment component that is needed for a group to coalesce. This is obvious in the physical patters of flocking or schooling, but also manifests in language (e.g. \enquote{political alignment}) \cite{denicola2017understanding} and fashion \cite{curran1999analysis}. 
	\item \textit{Influence} --  Agents within a specified range are capable of influencing each other's orientation and speed, inversely proportional to distance. This in turn influences heading, as more aligned agents have more time to influence each other \cite{olfati2007consensus}.
\end{enumerate}

There is considerable work in using groups of agents to evaluate fitness landscapes, where agents behave as particles, and increase their attractiveness based on their height in the landscape \cite{Kennedy2010}. This works well for difficult machine learning problems such as hyperparameter tuning \cite{ye2017particle}, but it does not capture the aspect of alignment in community behavior that is observed in sociological contexts. For our purposes, we chose the Reynolds model \cite{reynolds1987flocks}, which uses velocity and heading alignment as major components of its flocking and herding model. We modified the algorithm for high-dimensional belief spaces as we discuss in the methods section.

With respect to the individual agents, we can adjust what we term the \textit{social influence horizon} (SIH), or the area that the agent considers in its calculations. Closer agents have more influence over current agent's orientation and velocity. A low radius means that the agent has less social influence, which encourages exploration of the environment. The larger the radius becomes, the more the agent is dominated by social influence at the expense of environmental considerations.

The environment that these agents operate in represents the belief space that is mediated by technology. It supports variable, asymmetric visibility of one agent to another, boundary characteristics and the overall size and number of dimensions. In addition, it supports the storage of agent data at n-dimensional coordinates in the space. 

\subsection{Adversarial Herding}
Geils \cite{giles2016handbook} notes how social media can be used to increase polarization based on \textit{emergent} poles. In other words, \enquote{normal} opposing views can be amplified by attentive bad actors, with the goal of causing generalized  disruption. We added capabilities to the simulation that amplify the influence of selected individual agents to evaluate the effectiveness of potential disruptive mechanisms. Real world examples of this activity have been documented in the news media, where accounts of Russian troll farms organizing opposing groups \cite{stewart2018examining}. Most recently, Stella et al. have uncovered examples of bot-augmented human actors in the 2018 Italian general election \cite{stella2018influence}.

Though the term \enquote{Information Warfare} has been used in this context, we believe that a more appropriate metaphor is \textit{adversarial herding}. War describes conflict between two or more internally organized entities. Herding, on the other hand, can be defined as one entity causing many unwilling or unwitting entities to move in a desired direction \cite{strombom2014solving}. As with animals, herding can be performed directly, as with traditional methods using trained dogs to control livestock, or herding can be performed using subterfuge, as when native Americans would lure and drive buffalo over \textit{Pishkin cliffs} \cite{patent2006buffalo}. We believe that this \enquote{herding} approach more resembles what is known about social network exploitation. As such, we extended our base model to observe how that might manifest in a social media environment where it is easy to hide one's true identity\cite{kydd1997sheep}

Based on information collected from news stories and Gerasimov's work on nonmilitary methods in conflict development \cite{gerasimov2016value}, we added the ability to enable herding in the simulation, based on the following rules:

\begin{itemize}
	\item \textit{Herders can amplify arbitrary agents}, since they are not emotionally invested in their belief space position or orientation
	\item \textit{Herders appear like multiple individuals} (sockpuppets or sybils) that may seem close and trustworthy, but they are actually a distant monolithic entity that is aware of a much larger belief space.
	\item \textit{Herders amplify arbitrary pre-existing positions}. The insight is that \textit{they are not herding in a direction, but to increase polarization}
\end{itemize}

\section{Methods}

This study consisted of data generation and subsequent data analysis. We built a stand-alone simulator using Java that created the multidimensional belief environment and then populated it with agents. All agents are double buffered so that there are no sequential artifacts from agent interaction. The program can be run in interactive or batch mode. Sampled output was saved to Excel files.  

\subsection{Simulation}
The main components to be implemented were the agents and the environment they exist in. Since the number of beliefs that a person may hold is not limited by physical space, arbitrary numbers of dimensions need to accommodated. This was accomplished by collecting one-dimensional \textit{statements} into a structure defined as a \textit{belief}. Agents move through space based on Reynold's boids model\cite{reynolds1987flocks}, but with collision terms removed since individuals can hold identical views.

Each statement resembles a single element of an opinion dynamics model such as the ones used by Krause \cite{hegselmann2002opinion}.  For this work, each dimension was considered equivalent. Though this model is naive in that it is linear and orthogonal, social distance interactions across dissimilar spaces have been examined by Bogunia\cite{boguna2004models} and Schwammle\cite{schwammle2007different}. They show that our approach can be extended to handle non-linear spaces. 

\subsection{Position and orientation in high-dimensional belief spaces}
The agents' position vector of statements is updated by the orientation multiplied by the elapsed time since the last update (Equation \ref{eq:pos}):

\begin{equation} \label{eq:pos}
	\vec{acp} = \vec{app} + (\hat{aop})(vp)(dt)
\end{equation}

Where \begin{math} \hat{aop} \end{math} is the previous orientation unit vector, \begin{math} \vec{acp} \end{math} is the current position vector, \begin{math} \vec{app} \end{math} is the previous position vector, \textit{dt} is elapsed time, and \textit{vp} is the previous velocity of the agent.

The agent's target \enquote{orientation} vector of statements is updated by taking the average orientation of all agents that are within the SIH of the agent. This radius can be set to an arbitrary value. The influence drops off linearly until the radius is reached (Equation \ref{eq:distance} - notation from \cite{Iverson:1962}). The distinct phases identified in the results are exclusively the result of manipulating this radius.

\begin{equation} \label{eq:distance}
	\begin{aligned}
		\vec{to}_x=  {} 
		& \frac{\sum_{n=1}^{n = max [n \neq x]} \hat{aop}_n (1 - \frac{\| w_x\vec{app}_x - w_n\vec{app}_n \| }{r})}
		{1-\sum_{n=1}^{n=max [n \neq x]}\|w_x\vec{app}_n - w_n\vec{app}_x\| } \\ 
		& [\|w_x\vec{app}_x - w_n\vec{app}_n \| < w_nr]
	\end{aligned}
\end{equation}

Where \begin{math}	\vec{to}\end{math} is the target orientation vector,
\begin{math}\hat{aop}\end{math} is the previous orientation unit vector,
\begin{math}\vec{app}\end{math} is the previous position, and  
\textit{r} is the social influence horizon. The \textit{w} term is the scalar weight of the agent's influence, which is set to 1.0 in non-herding cases.

\begin{figure}[h]
	\centering
	\fbox{\pdftooltip{\includegraphics[height=10em]{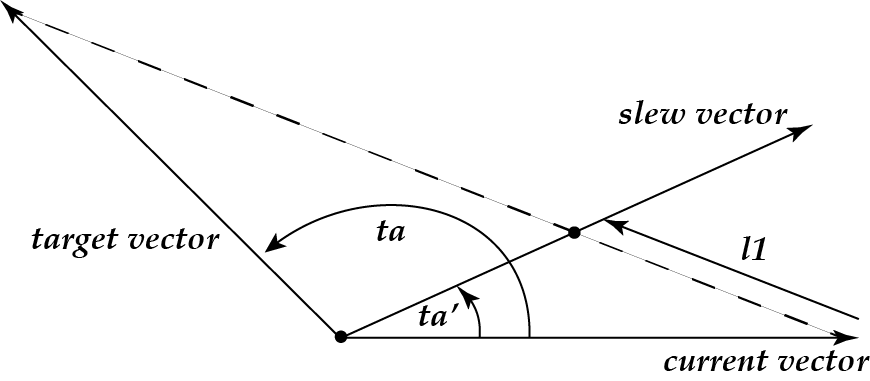}}{Calculation of projected slew vector in arbitrary dimensions. The \enquote{transition} vector between the current agent orientation and the target orientation is calculated, then the maximum turning angle is projected on that plane. This determines the amount to move along the transition vector, which produces a new agent orientation vector.}}
	\caption{\label{fig:slewAngle} Calculation of projected \enquote{slew} vector}
\end{figure}

The agent then interpolates its orientation towards the goal vector as a function of time (figure \ref{fig:slewAngle}). First, the angle between the current and target orientation is calculated as a direction cosine. This allows for 2D calculations in the plane of rotation (Equation \ref{eq:interpolate}):

\begin{equation} \label{eq:interpolate}
ta = cos^{-1} (\frac{\hat{aop} \cdot \vec{to}}{\|\vec{to}\|})
\end{equation}

Where
\textit{ta} is the 2D angle, 
\begin{math}\hat{aop}\end{math} is the previous orientation unit vector, and 
\begin{math}\vec{to}\end{math} is the target orientation vector

The maximum incremental angle that the agent can rotate through (\textit{ta'}) is calculated as a fraction of the above, constrained by the turn rate of the agent and elapsed time since the last simulation step (Equation \ref{eq:slew}):

\begin{equation} \label{eq:slew}
ta' = (ta)(rate)(dt)
\end{equation}

The line formed by the current vector and the target vector can then be intersected with the slew vector. This intersection allows us to calculate the length of the vector that we add to the current \textit{n}-dimensional orientation vector to produce the new orientation. Lastly, the scaled vector is added to the agent vector and normalized.

\begin{figure*}
	\centering
	\begin{minipage}{.33\textwidth}
		\centering
		\fbox{\pdftooltip{\includegraphics[height=12em]{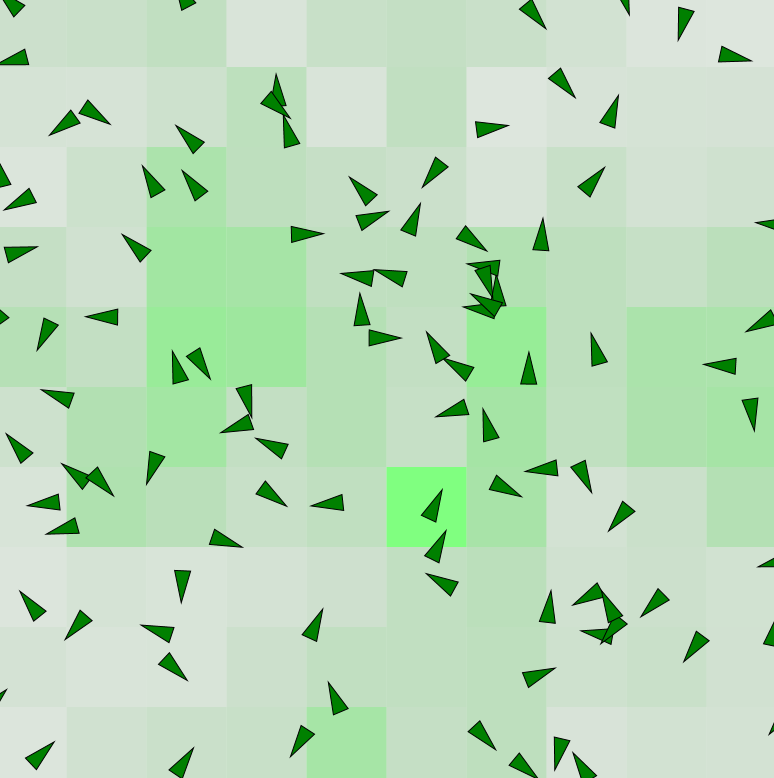}}{Nomadic Phase. Agents are evenly distributed through the environment, which is reflected in the underlying heatmap.}}
		\caption{\label{fig:explorer} Nomadic Phase}
	\end{minipage}%
	\begin{minipage}{.33\textwidth}
		\centering
		\fbox{\pdftooltip{\includegraphics[height=12em]{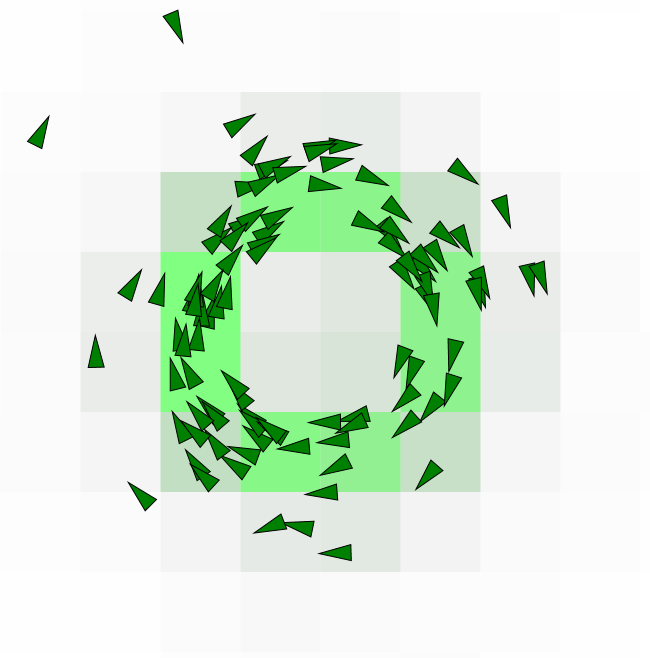}}{Flocking Phase. Agents \enquote{orbit} the center of the simulation as they attempt to align speed and orientation with their nearby neighbors.}}
		\caption{\label{fig:coloredFlocking} Flocking Phase}
	\end{minipage}%
	\begin{minipage}{.33\textwidth}
		\centering
		\fbox{\pdftooltip{\includegraphics[height=12em]{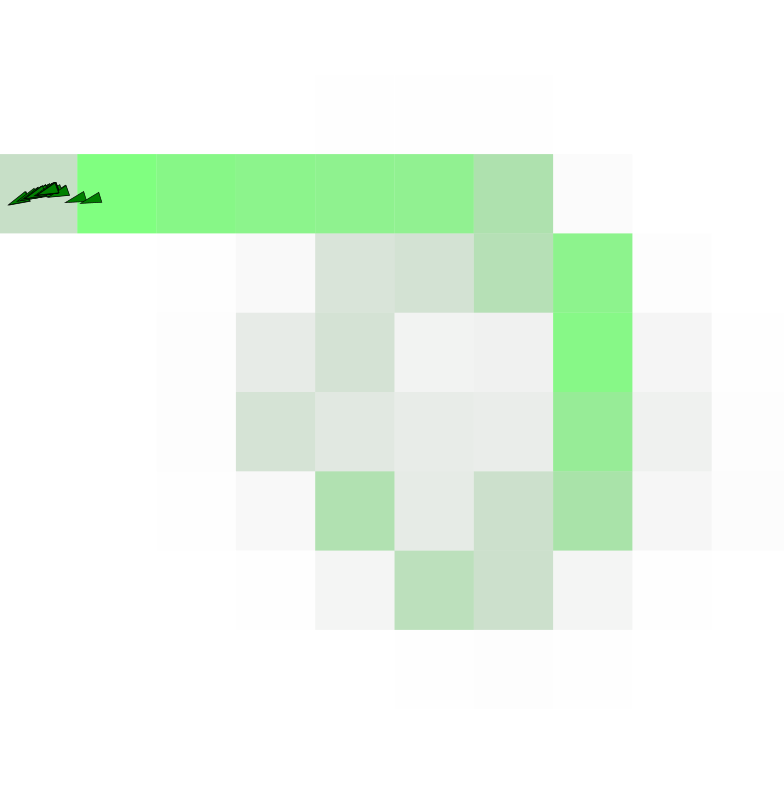}}{Stampede Phase. Agents cluster tightly together. Once they substantially match position, velocity and orientation there is no external influence to return them to the center of the belief space, so they head towards the borders in a stampede.}}
		\caption{\label{fig:stampede} Stampede Phase}
	\end{minipage}%
\end{figure*}

This model produces distinct emergent patterns, by adjusting only the SIH:: 

\textit{Nomadic Phase} (figure \ref{fig:explorer}) - A low SIH means low influence by other agents, so each agent moves in their own direction engaging in the \enquote{explore} phase of the Multi-Armed Bandit problem.

\textit{Flocking Phase} (figure \ref{fig:coloredFlocking}) - An intermediate SIH results in an agent whose movement is affected by nearby individuals. There is alignment with neighbors, but there is sufficient diversity so that orientations change over time

\textit{Stampede Phase} (figure \ref{fig:stampede}) - At high SIH, all members are exposed equally to each other in what is essentially a fully connected graph, which synchronize easily \cite{olfati2007consensus}. As such, alignment can become total and supports runaway \cite{lande1981models} conditions. In the Multi-Armed Bandit problem this would be \enquote{exploit} without any explore. This is also represented in other literature as \enquote{filter bubbles}, \enquote{echo chambers} \cite{flaxman2016filter}, \enquote{group polarization}, and \enquote{extremism} \cite{moscovici1994conflict}.

To support experimentation in this simulated domain, the simulation was configured so that the following variables could be manipulated.
\begin{itemize}
	\item Number of agents
	\item One or two populations, with separate social influence horizons (0 - 10 units),  operating independently or aware of each other
	\item Data accumulation in the environment by \textit{n}-dimensional cell. Heatmaps, etc.
	\item Number of dimensions (2 - 10)
	\item Environment size (0 - 10 units)
	\item Environment border:
	\begin{itemize}
		\item NONE: No effect when reaching the limit of the environment
		\item REFLECT: Agents are reflected back whenever they cross a border
		\item RESPAWN: If a border is crossed, the agent is re-initialized in the environment with a random heading, position and speed.
	\end{itemize}
\end{itemize}

\subsection{Adversarial Herding}
To create inputs that resemble hypothetical herding behaviors, and to be able to view them, the model was extended to accommodate the following additional capabilities:

\begin{itemize}
	\item The selected agent's weight and social influence are increased to \textit{w'} and \textit{r'}. \textit{w'} represents amplification by trolls, bots, etc. A large \textit{r'} means that the bots can swamp other, normally \enquote{closer} signals. This models the effect of a monolithic entity controlling thousands of bots across the belief space \cite{stella2018influence}.
	\item There are three optional herding modes: 
	\begin{itemize}
		\item Where the agent closest to the average heading is amplified. This mimics the effect of bots retweeting human actors that align with the adversaries' goals.
		\item A random \enquote{leader} is chosen for the duration of the simulation. This mimics the effect of sustained support of a single individual over time by bots and sybils.
		\item  A randomly chosen agent is amplified each cycle. This does not mimic any known technique, but was chosen because it is a simple case that randomly contaminates the social influence radius that is used to calculate group interaction.
	\end{itemize}
\end{itemize}

\subsection{Finding Patterns}
Though the DISI patterns described above were visible and apparent when using the system interactively, the goal is to be able to detect the patterns programmatically. Our starting point is the work of Boguna et.al \cite{boguna2004models}, who developed a set of models based on a mathematical abstraction of \enquote{social distance}. We used Dynamic Time Warping (DTW) \cite{salvador2007toward} to determine the overall social distance between paths of individual agents. DTW works by determining how to transform one sequence into optimal alignment with another using non-linear mapping. This is different from linear transformations such as Least Squares, which produce an approximate fit. As outputs, DTW produces a \enquote{warping path} and a total \enquote{warping distance}. This approach was sufficient to discriminate between the populations, while being fast enough to be used for large-scale analytics.

\section{Results}
All experiments were managed by configuration files for repeatability, with typical experiments consisting of 100 agents run 10 times through each set of conditions. Initial experiments were done to determine if the number of dimensions altered agent behaviors. We found that the SIH had to be multiplied by the square root of the number of dimensions to produce the same agent behaviors. This is an example of the \enquote{curse of dimensionality} \cite{bellman2013dynamic}, which refers to the difficulty of calculating meaningful Euclidean distance in high-dimensional space. This may explain why polarization only happens after concepts have been simplified. Open discussion can be a form of dimension reduction \cite{moscovici1994conflict} which creates a common framing that supports consensus \cite{bacharach2006beyond}. Based on this result, the majority of simulations were run in two dimensions which eliminates projection artifacts from the visualizations.

Regardless of the number of dimensions (2 - 10 tested), we were able to see that agent behavior rapidly manifested in three phases by varying only the SIH. These three phases can be seen in Figures \ref{fig:explorer}, \ref{fig:coloredFlocking} and \ref{fig:stampede}.

\begin{figure}[h]
	\centering
	\begin{minipage}{.25\textwidth}
		\centering
		\fbox{\pdftooltip{\includegraphics[height=9.5em]{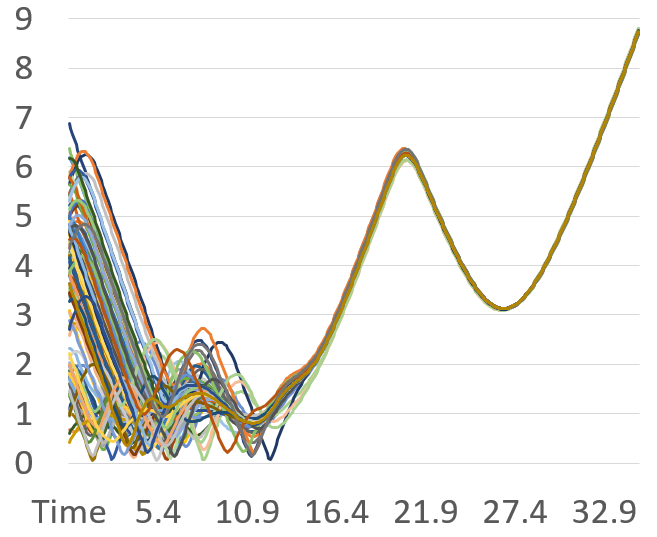}}{Stampede timeline of agent distance from the origin. Agents converge from their original random distribution and start an elliptical spiral. Once sufficiently compact, the cluster moves towards the boundary.}}
		\caption{\label{fig:RunawayTrace} Stampede distance}
	\end{minipage}%
	\begin{minipage}{.25\textwidth}
		\centering
		\fbox{\pdftooltip{\includegraphics[height=9.5em]{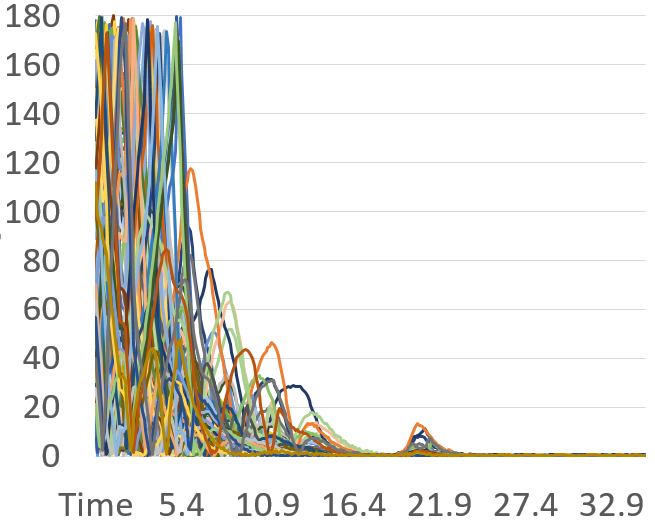}}{Stampede timeline of agent heading with respect to the average. Agents converge from their original random headings. With one exception that tracks to the turn at 21.9 in figure \ref{fig:RunawayTrace}, the headings tightly converge and remain in the \enquote{stampede condition}.}}
		\caption{\label{fig:AvgCenterConverge} Stampede heading}
	\end{minipage}%
\end{figure}

For these simulation runs, agents were initialized on a range of (-5.0, 5.0) on each dimension. A reflective barrier was placed at (-10.0, 10.0). This embodies the intuition that many concepts have inherent limits. For example, in fashion, a skirt has practical limits in length and width\cite{curran1999analysis}. 

The first phase is determined entirely by the initial random generation of the agents. They continue along their paths until they encounter the containing barrier. The behavior is random with no emergent pattern. The second phase is the richest, characterized by the emergence of \textit{flocks} or \textit{schools}. The third phase represents an example of a runaway polarization condition or \textit{stampede}. Figure \ref{fig:stampede} shows a tightly clustered group heading towards the edge of the environment. Figures \ref{fig:RunawayTrace} shows the convergence of the agents as they cluster with respect to the belief space origin. Figure \ref{fig:AvgCenterConverge} shows the same event with respect to heading.  All agents become tightly aligned and clustered, and their position in space becomes more extreme over time. The only thing preventing the polarized group from heading off into infinity is the boundary.

\subsection{Emergent Group Behavior}

\begin{figure}[h]
	\centering
	\fbox{\pdftooltip{\includegraphics[width=24em]{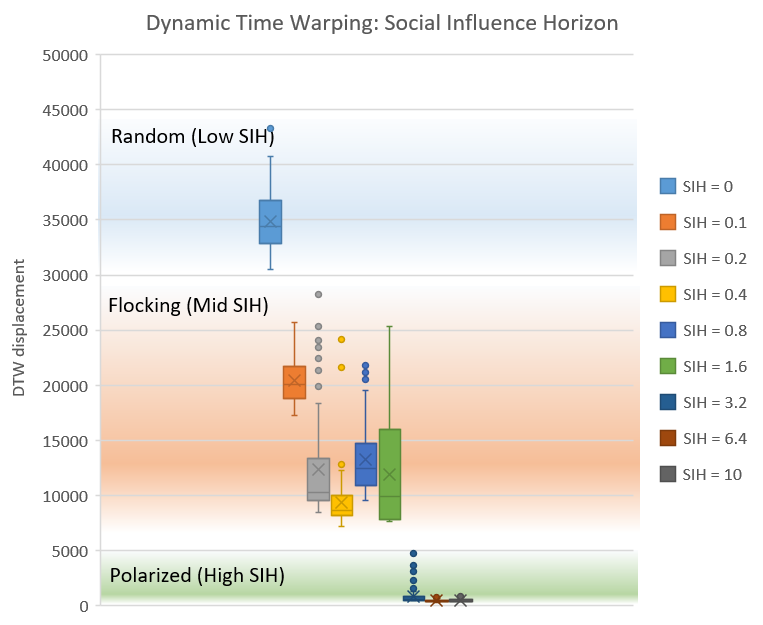}}{Nomad, Flocking and Stampede dynamic time warping. The box plots show three distinct populations. For zero SIH (random behavior), the DTW is the highest, between 30,000 and 42,000. For SIH of 0.1 through 1.6 (flocking behavior), the plots are clustered between 7,000 and 28,000. For SIH of 3.2 and above (stampede behavior) the plots are tightly clustered between 0 and 5,000.}}
	\caption{\label{fig:populations} Nomad, Flocking and Stampede DTW}
\end{figure}

We used Dynamic Time Warping (DTW) as discussed in the methods section and implemented in Java--ML to determine population membership with respect to SIH. DTW attempts to find the lowest distance that one set of points need to be moved to exactly match another sequence of points. We then build a matrix for the DTW distance between each agent and sum each column to compute the overall distance of one agent to the other agents in the simulation. The distribution of DTW distance by agent SIH is shown in Figure \ref{fig:populations}.  The phase changes (nomad, flocking, stampede) are distinctive and \textit{non-overlapping} in our datasets.

This strongly supports the observation that there are three distinct phases of behavior in the agents that are affected by the size of the SIH. Further, these phases are sufficiently distinct that they can be statistically recognized.

\subsection{Interactions between populations}
\begin{figure*}
	\centering
	\begin{minipage}{.33\textwidth}
		\centering
		\fbox{\pdftooltip{\includegraphics[height=12em]{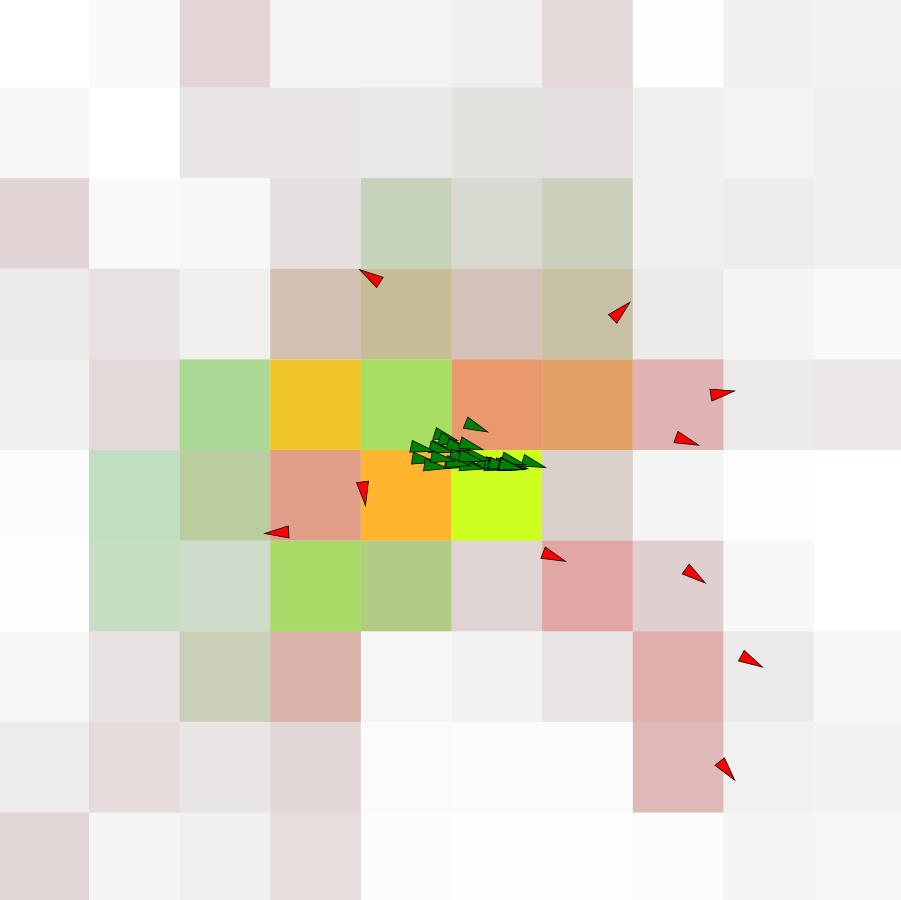}}{Nomadic agents influencing stampeding agents towards the center of the simulation}}
		\caption{\label{fig:Explore-Exploit} Nomad influence}
	\end{minipage}%
	\begin{minipage}{.33\textwidth}
		\centering
		\fbox{\pdftooltip{\includegraphics[height=12em]{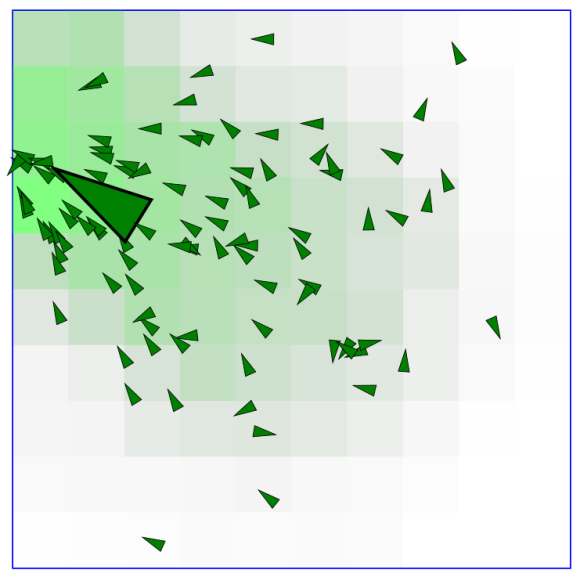}}{Single amplified agent influencing self-sustaining stampede towards lethal boundary}}
		\caption{\label{fig:PishkinEffect} Pishkin Effect}
	\end{minipage}%
	\begin{minipage}{.33\textwidth}
		\centering
		\fbox{\pdftooltip{\includegraphics[height = 12em]{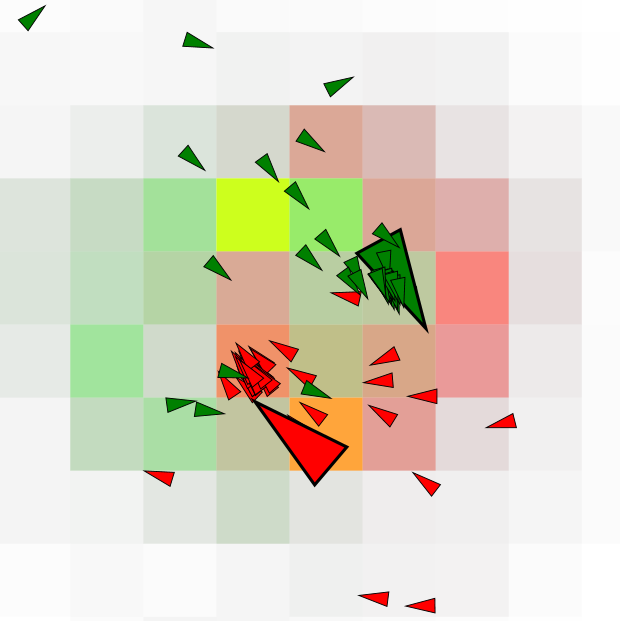}}{Two amplified agents herding populations in opposite belief directions}}
		\caption{\label{fig:adversarialHerding}Opposing Herding}
	\end{minipage}%
\end{figure*}
We also examined the interactions between two populations, each with a different SIH. Multiple studies across different disciplines ranging from neurology\cite{cohen2007should} to computer-human interaction\cite{munson2010presenting} have shown that populations often have explorer and exploiter subgroups. In nature, many effective strategies revolve around a majority exploit/minority explore pattern\cite{cohen2007should}.

In one set of simulations, 10\% of the population were given zero SIH, letting them explore the environment uninfluenced, while the other 90\% were given the highest SIH (the size of the environment), which in prior runs had resulted in the group polarization of figure \ref{fig:stampede}. These percentages reflect the results found by Cohen\cite{cohen2007should} as well as the percentage of diverse news consumers found by Flaxman et. al. in their study of browser logs\cite{flaxman2016filter}.

The results of mixing these populations was startling. Although still tightly clustered, the stampede group would rarely encounter the simulation boundary and would instead be influenced towards the center by the presence of nomads. Although these agents were still a polarized group, they were no longer in a runaway condition (Figure \ref{fig:Explore-Exploit}). 

The reason that the \textit{nomads} are so successful in adjusting the trajectory of polarized groups has to do with their distribution. Because nomads maintain their random orientations throughout the simulation, they effectively provide a normal probability distribution over the environment within each dimension. What this model implies is that a sufficiently diverse population covers an belief space in such a way that if their \textit{position} is visible to another population, it can have the effect of providing an attraction to the center of the environment. 

\subsection{Adversarial Herding}

\begin{figure}
	\centering
	\fbox{\pdftooltip{\includegraphics[width=24em]{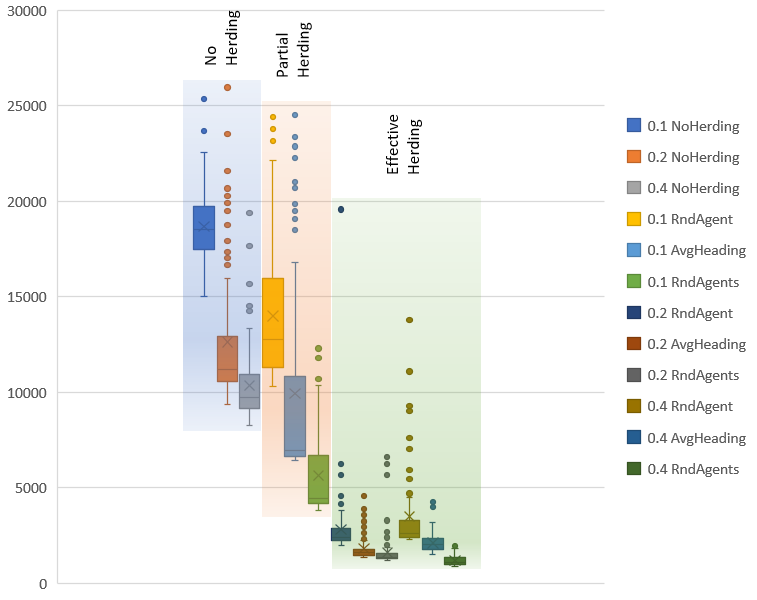}}{Dynamic time warping values for herding. Compared with the control of  \enquote{no herding} for SIH of 0.1 - 0.4 where there is no stampeding, herding results in more clustered activity for SIH of 0.1, and tightly clustered \enquote{stampedes} at SIH of 0.2 and 0.4.}}
	\caption{\label{fig:HerdingImpact} No, Partial, and Effective Herding DTW}
\end{figure}

Adding arbitrary amplification to a single \enquote{super agent} as a way of creating a Pishkin-style stampede appears to be very effective. The impact of herding on DTW measures is shown in figure \ref{fig:HerdingImpact}. The \enquote{No Herding} populations are the same populations as shown in figure \ref{fig:populations} at the low end of the \enquote{flocking} SIH (the green middle band). The middle \enquote{Partial Herding} depicts the herding algorithm on the 0.1 unit influence horizon. From left to right, the herding implementation is:

\begin{enumerate}
	\item Single, randomly chosen agent
	\item The agent currently closest to the average flock heading
	\item A random agent chosen at every sample
\end{enumerate}

The least effective strategy is to choose a single agent. Somewhat counterintuitively, the most effective polarizing strategy is to choose random agents and amplify them for a short period. This between-strategies relationship is maintained in the green \enquote{Effective Herding} region, though the overall clustering is tighter and the variance between the results is much lower.

The reason for this is that randomly iterating over the entire population gives the entire population a high \enquote{effective} SIH. In the example shown in figure \ref{fig:populations}, the average DTW for the maximum SIH population evaluated is approximately half of the lowest population in figure \ref{fig:HerdingImpact}. This seems reasonable, since the \enquote{natural} case is where all the agents have identical horizons should converge faster than any process that depends on a single agent, no matter how effectively distributed.  

When a single, random agent is chosen, the stampede effect ends soon after the super agent encounters a lethal condition (The RESPAWN boundary in these simulations). Once free of the agent's amplified influence, the other agents settle back into their normal flocking behavior. In the case where random agents are amplified, the results are more complicated. In the case where direction is being most influenced by completely random agents, the tendency towards polarization is maintained in the population, though it is unfocused, and takes a while to build back to the runaway condition. In the case where the super agent is chosen that most matches the average heading of the group, a more stable condition can arise where the clustering, orientation and velocity of the stampede is maintained independently of the particular agents involved. As stampeding agents encounter the lethal condition, they are continuously replaced by new agents that are randomly initialized into the environment to replace the terminated ones, but under the influence of the current super agent, these new agents are influenced by the ongoing stampede pattern and head back to the \enquote{cliff}. 

To see how opposing polarized groups might be created, we also implemented a variant of herding implementation(2), where an agent in a separate population that most matches the inverse vector of the first population is amplified (figure \ref{fig:adversarialHerding}). This mimics the effect of an adversary supporting antagonistic extremism between groups. This resembles the RU-IRA interferences with the \#blacklivesmatter/\#bluelivesmatter twitter interaction \cite{stewart2018examining}.

\section{Discussion}
We have built a framework, based on a Reynolds \cite{reynolds1987flocks} agent-based simulation to explore the emergent group behaviors based on interactions between individuals' heading and orientation in a belief space hypercube. This simulation is a starting place to generate identifiable behaviors in belief space. We have begun work to look for similar patterns in the data produced by human users in subsequent studies. As our understanding of navigation in belief space improves, the model will be updated with more sophisticated rules grounded in observed human behavior patterns. 

Animals make decisions as groups that manifest as a variety of patterns, including swarms, flocks, schools, and stampedes. The inadvertent social information provided by these groups allow other animals to infer ecosystem qualities. Humans appear to traverse belief space analogous to how animals move in physical space. Heading, velocity, and influence distance are qualities that we find are important to this process and occur in a wide variety of contexts. Recent support for this intuition have come from studies of group neural alignment by Stephens \cite{stephens2010speaker}, Gallotti \cite{gallotti2017alignment} and others. In simulation, the patterns of \textit{digital inadvertent social information} (DISI) can be detected efficiently and on large data sets using DTW. Since analyzing DISI does not depend on language-dependent text analytics, this approach should be fundamentally domain independent. 

The number of dimensions seems to matter. Low dimensions make it easier to initiate a stampede. Moscovici and Diose showed that unconstrained group deliberation will eventually reduce the subject of the discussion to a simplified representation, within which the group can polarize around a consensus. When structures are added to restrict this simplification process, group deliberation often leads to compromise, rather than polarization\cite{moscovici1994conflict}. 

Individuals and groups must make decisions with incomplete information. Using social cues as a proxy for direct evaluation of a problem has the benefit of less computation and shared risk, but contains the threat of groupthink. How this interaction plays out affects individual and group behavior in a dynamic, emergent way. Biological systems appear to have evolved the explore/exploit behavior pattern to address this issue. In a population, some relatively small percentage is biased towards nomadic exploration, while the majority tend to exploit their situation. This makes evolutionary sense, as exploration is risky. However, if a catastrophe strikes the main population, the species can rebuild from the nomad diaspora.

Stampedes are highly dynamic but fundamentally simple systems where individual environmental interaction is overridden by a single social reality\cite{granovetter1978threshold}. Once established stampedes often continue until they encounter a sufficient disrupting condition. At a societal level in humans, stampedes, or echo chambers can be an existential threat to ongoing governmental and cultural norms \cite{sunstein2002law, arendt1973origins}. Increased awareness of a small population of nomadic explorers can positively influence the stable equilibrium of a stampede.

\section{Implications for Design}
We are influenced by psychological rules we can't control. For example, the title of this paper plays with our need to know \enquote{hidden knowledge}. Similarly, the gaps that appear as we apply our understanding of physical spaces to virtual domains are inherently dangerous and exploitable. 

The simple trick is this: \textit{Changing the awareness of others with different levels of alignment disrupts communities}. This can be positive or negative. It can drive a polarizing group towards extremism, or hinder a stampeding mob.

The reason that this trick works is that when technology mediates communication, it changes the social cues that let us determine the trustworthiness of the information we receive. When we transfer our understanding of trustworthiness from the richness of the physical world to the sparse abstraction of online environments, the boundaries between well-supported evidence and naive belief become obscured. These low-dimensional spaces can be breeding grounds for misinformation.  

As we have seen with the Jade Helm conspiracy theory \cite{quattrociocchi2017inside}, these \enquote{stampedes} can be naturally occurring. However, they can also be accentuated by an adversary that simply \textit{amplifies} credible voices in the community, increasing the reach of the misinformation and bringing the credulous into alignment.

Poor information at an individual level that leads to Echo chambers are a problem. Echo chambers at a societal level can be an existential threat to ongoing governmental and cultural norms. Computer-mediated communication is very much a personalized interaction, but with effects that scale to populations. What changes could be made to these individual interactions to have desired socio-organizational and cultural effects? 

Our research has indicated that an awareness of nomadic/explorer activity in belief space may help nudge stampeding groups away from a terminal trajectory and back towards \enquote{average} beliefs. Tajfel \cite{tajfel2004social} states that groups can exist \enquote{in opposition}, so providing counter-narratives may be ineffective. Rather, we think that a potential approach to reducing online polarization is to inject  diversity into user’s social media, search results, and video feeds  \cite{andre2009discovery}.  The infrastructure exists for this already in platform's support of advertising. An existing template could be the Public Service Announcement (PSA).

US Broadcasters since 1927, have been obligated to \enquote{serve the public interest} in exchange for spectrum rights. One way that this has been addressed is through the creation of the PSA, \enquote{the purpose of which is to improve the health, safety, welfare, or enhancement of people’s lives and the more effective and beneficial functioning of their community, state or region} \cite{lamay2002public}.

We believe that PSAs can be repurposed to support diversity injection (DI) through the following:

\begin{enumerate}
	\item Random, non-political content designed to expand information horizons, analogous to clicking the \enquote{random article} link on Wikipedia.
	\item Progressive levels of detail starting with an informative \enquote{hook} presented in social feeds or search results. Users should be able to explore as much or little as they want.
	\item Simultaneous presentation to large populations. Google has been approximating this with their \enquote{doodle} since 1998, with widespread positive feedback, which indicates that there may be good receptivity to common serendipitous information.
	\item Format should reflect the medium, i.e. text, images and videos.
	\item Content should be easily verifiable, recognizable, and difficult to spoof.
\end{enumerate}

We believe that such DI mechanisms as described above can serve as a \enquote{first do no harm} initial step in addressing the current crisis of misinformation. By nudging users towards an increased awareness of a wider world, DI interferes with the processes that lead to belief stampedes by increasing the number of dimensions, and awareness of different paths that others are taking. 

As we gain deeper understanding of the mechanisms that influence group behaviors, it may be possible to further refine our designs and interfaces so that they no longer promote extremism at the socio-organizational level while still providing value at individual and small group levels. Generally, designs need to take into account how individual interaction manifests at different social scales. We believe that many of the current issues that are plaguing the largest-scale information providers, such as Facebook and Google derive from a failure to design for large scale belief behavior, as implemented in software and business models. 

\section{Conclusions}
The benefits of cooperative structures appear continuously in evolution. Examples abound from cellular organelles to multicellular organisms to flocks, schools, forests and even ecosystems \cite{west2007evolutionary}. In these natural systems, mechanisms ranging from scent, to sight to sound evolved to provide relevant and appropriate information for the collective to make optimal decisions \cite{dall2005information}. An example of this is a school of fish able to find food when the scent is faint and patchy \cite{grunbaum1998schooling}. Though generally effective, collective approaches can fail spectacularly. Even with co-evolved information and social systems, buffalo stampede, whales beach, and locusts swarm into plagues. 

Emergent structures for human social evolution have also developed, from tribe through village to city, nation-state and pan-national movement. As our social structures increased in size and sophistication, technology has provided us tools with an ever-increasing power to manipulate, store and transfer information. But our biological mechanisms for dealing with what we experience through our screens hasn't changed much since our ancestors were making tools from flint and obsidian. Approaches that worked when we were members of small tribes hunting and gathering can easily be fooled when the sources of our information are obscured, either by technological and economic constraints or with more malevolent intent. For cooperative structures to function, the quality of information provided by the members of the group must be judged with respect to the source's trustworthiness, not just its credibility. When the information flow between agents is distorted or misrepresented, \enquote{belief stampedes} can result. Examples include religious cults, market bubbles and crashes, and political phenomena like totalitarianism. 

In addition to animal and human behavior, it could be wise to consider how these widespread and generalizable relationships might affect intelligent machines. The \enquote{flash crash} of 2010 could be regarded as a stampede of trading programs \cite{kirilenko2017flash}. Given the rise of robotic systems such as self driving cars, it could be wise to consider the ramifications of these tendencies in populations of autonomous vehicles.

Recently, it has become popular to revisit moral dilemmas such as the Trolley Problem \cite{thomson1976killing} in the light of progress with intelligent machines \cite{waser2008discovering} \cite{deng2015robot}. Does a self-driving car decide to crash, risking its occupant to save the life of a child that it sees in its path? Although the authors certainly believe that this thinking is valuable, we also feel it is important to think about such problems when they occur at scale. Where does the moral responsibility lie? With the individual machine? Or the emergent, collective intelligence of the interlinked system?

Runaway group behavior can emerge if the conditions are right. Just as a canyon reduces the options for a herd of buffalo, making it easier for them to be spooked into stampeding, so too a reduction in the options available to intelligent systems can have profound effects. In the massive fires that swept through greater Los Angeles in December of 2017, police determined that GPS systems were attempting to re-route users around the closed highway 405 through burning neighborhoods \cite{nelson_2017}. It doesn't take much imagination to extrapolate what could happen if instead of human operators, autonomous vehicles with no training on how to behave in such massive disasters were the primary form of transportation. Without a pre-trained awareness of the dangers of flames and heat, large numbers of vehicles would obediently follow their carefully optimized route into the fire. As they enter the inferno, their sensors begin to malfunction. The antennas that transmit position burn off and the cars become invisible to the network so the path continues to classify as clear and open. Vehicles by the hundreds continue to flow towards this virtual Pishkin, as betrayed by their design as buffalo were betrayed by their instincts hundreds of years before.


\end{document}